\documentclass[9pt]{article}
\usepackage{spconf,amsmath,graphicx}
\usepackage{cite}
\usepackage{bibspacing}
\usepackage{booktabs}
\usepackage{color}
\usepackage{amsfonts}
\usepackage{multirow}
\usepackage{amssymb}
\usepackage{array}
\usepackage{authblk}
\usepackage{threeparttable}
\newcommand{\PreserveBackslash}[1]{\let\temp=\\#1\let\\=\temp}
\newcolumntype{C}[1]{>{\PreserveBackslash\centering}p{#1}}
\newcolumntype{R}[1]{>{\PreserveBackslash\raggedleft}p{#1}}
\newcolumntype{L}[1]{>{\PreserveBackslash\raggedright}p{#1}}
\usepackage{enumitem}

\usepackage[hang]{footmisc}

\AtBeginDocument{
 \footnotesize\setlength{\footnotemargin}{0pt}\normalsize
 \edef\hangfootparindent{\the\parindent}
}

\usepackage{lipsum} % mock text

\makeatletter
\newcommand\email[2][]%
   {\newaffiltrue\let\AB@blk@and\AB@pand
      \if\relax#1\relax\def\AB@note{\AB@thenote}\else\def\AB@note{\relax}%
        \setcounter{Maxaffil}{0}\fi
      \begingroup
        \let\protect\@unexpandable@protect
        \def\thanks{\protect\thanks}\def\footnote{\protect\footnote}%
        \@temptokena=\expandafter{\AB@authors}%
        {\def\\{\protect\\\protect\Affilfont}\xdef\AB@temp{#2}}%
         \xdef\AB@authors{\the\@temptokena\AB@las\AB@au@str
         \protect\\[\affilsep]\protect\Affilfont\AB@temp}%
         \gdef\AB@las{}\gdef\AB@au@str{}%
        {\def\\{, \ignorespaces}\xdef\AB@temp{#2}}%
        \@temptokena=\expandafter{\AB@affillist}%
        \xdef\AB@affillist{\the\@temptokena \AB@affilsep
          \AB@affilnote{}\protect\Affilfont\AB@temp}%
      \endgroup
       \let\AB@affilsep\AB@affilsepx
}
\makeatother

\title{HYBRIDFORMER: IMPROVING SQUEEZEFORMER WITH HYBRID ATTENTION

AND NSR MECHANISM}
\author[1*]{Yuguang Yang}
\author[2*]{Yu Pan}
\author[1]{Jingjing Yin}
\author[1]{Jiangyu Han}
\author[2$\dagger$]{Lei Ma}
\author[1$\dagger$]{Heng Lu}
\affil[1]{Ximalaya Inc., ShangHai, China}
\affil[2]{University of Alberta, Edmonton, Canada}
\email{\{yuguang.yang, jingjing.yin, jiangyu.han, bear.lu\}@ximalaya.com, yp14@ualberta.ca, ma.lei@acm.org}
\begin{document}
%\ninept
%
\maketitle
\begin{abstract}
SqueezeFormer has recently shown impressive performance in automatic speech recognition (ASR). However, its inference speed suffers from the quadratic complexity of softmax-attention (SA). In addition, limited by the large convolution kernel size, the local modeling ability of SqueezeFormer is insufficient. In this paper, we propose a novel method HybridFormer to improve SqueezeFormer in a fast and efficient way. Specifically, we first incorporate linear attention (LA) and propose a hybrid LASA paradigm to increase the model's inference speed. Second, a hybrid neural architecture search (NAS) guided structural re-parameterization (SRep) mechanism, termed NSR, is proposed to enhance the ability of the model to extract local interactions.
Extensive experiments conducted on the LibriSpeech dataset demonstrate that our proposed HybridFormer can achieve a 9.1\% relative word error rate (WER) reduction over SqueezeFormer on the test-other dataset. Furthermore, when input speech is 30s, the HybridFormer can improve the model's inference speed up to 18\%. Our source code is available online$^{1}$.

\end{abstract}

\vspace{-3.5mm}
\section{Introduction}
\label{sec:intro}%
{
\let\thefootnote\relax
\footnote{*Yu Pan and Yuguang Yang are co-first authors and contribute equally. }
\footnote{$\dagger$ Lei Ma and Heng Lu are the corresponding author.}
\footnote{$^{1}$ Code: https://github.com/yygle/HybridFormer}
}

\vspace{-5.5mm}
\noindent 
Over the past years, end-to-end (E2E) based speech recognition has become mainstream for ASR society. 
Among them, the transformer-based methods \cite{num1vaswani2017attention,num3zhang2020transformer,num4miao2020transformer,num6gao2022paraformer} are adept at exploiting long-term interactions, while convolution neural network (CNN) based methods \cite{num7li2019jasper,num8kriman2020quartznet,num10majumdar2021citrinet} perform well in extracting local context-dependent features. Afterward, by incorporating CNN into transformer, Conformer \cite{num11gulati2020conformer} is proposed to capture both global and local acoustic information at the same time.

Although Conformer has shown promising performance in various speech related tasks \cite{num12li2022ead,num13narayanan2021cross,num14chen2021continuous}, authors of SqueezeFormer \cite{num21kim2022squeezeformer} argued that the design choices of Conformer are sub-optimal and there is great unnecessary temporal redundancy in the deep layers of the model. Therefore, as shown in the left part of Fig.1, SqueezeFormer refines the micro and macro architectures of Conformer to improve the ASR performance. Besides,
by computing attention over a sequence with reduced length, its computational overhead 
can be further decreased.

% Recently, \cite{num21kim2022squeezeformer} proposed a novel method SqueezeFormer and achieved state-of-the-art performance. Through systematical reexamination, SqueezeFormer found that the design choices of Conformer are not quite optimal, and there is great unnecessary temporal redundancy in the deeper layers of the model. Thus, as depicted in the left part of Fig. 1, SqueezeFormer refined the micro and macro architectures of Conformer to improve the model's performance and decreased the computational overhead by computing attention over a reduced sequence length in the squeeze blocks of its model (i.e., when the sampling rate is 80ms).

% Recently, SqueezeFormer \cite{num21kim2022squeezeformer} found that the design choices of Conformer are not quite optimal, and there is great unnecessary temporal redundancy in the deeper layers of the model. Thus, after systematical reexamination, SqueezeFormer refined the micro and macro architectures of Conformer to improve the model's performance and decreased the computational overhead by computing attention over a reduced sequence length in the squeeze blocks of its model (i.e., when the sampling rate is 80ms), which is depicted in the left part of Fig. 1.

% \vspace{-2mm}

Nevertheless, in this paper, we believe that SqueezeFormer can be improved in several aspects. First, like other SA-based ASR models, SqueezeFormer also suffers from the quadratic temporal-space complexity concerning the input sequence length, especially in the unsqueeze blocks. Second, although SqueezeFormer is a hybrid attention-convolution model like Conformer, constrained by the large convolution kernel size (i.e., 31), its ability to exploit local features is still relatively limited.

% \vspace{-2mm}
Motivated by the aforementioned analysis, we propose a fast and efficient method termed HybridFormer to bridge this gap. First, we propose a novel hybrid \textbf{L}inear \textbf{A}ttention\cite{num15tay2020efficient,num16su2021roformer,num17wang2020linformer,num18huang2019interlaced,num19qin2022cosformer,num20choromanski2020rethinking} and \textbf{S}oftmax \textbf{A}ttention (LASA) paradigm, based on SqueezeFormer's architecture, to alleviate the quadratic complexity of the SA operation.
% using LA\cite{} while retaining high performance.
% Second, we present a novel hybrid NAS-guided SRep mechanism which is called NSR and apply it to the convolution module to improve the performance of the model while not increasing any computational complexity during the inference phase. 
Second, we present a novel hybrid \textbf{N}AS-guided \textbf{SR}ep mechanism (NSR) and apply it to the convolution module to improve system performance without increasing any computational complexity during inference.
Third, we contribute the first AED\cite{aed} based SqueezeFormer on WeNet \cite{num22wu2021u2++} toolkit and the core code of our proposed HybridFormer is also available online.
% Third, we contribute the first AED\cite{}-based SqueezeFormer to WeNet \cite{num22wu2021u2++} toolkit and release our source code on Github.

The remainder of this paper is organized as follows. In Section \ref{sec:METHODOLOGY}, we describe the details of the proposed method HybridFormer. In Section 3, the experimental dataset and setups are introduced to verify the effectiveness of the proposed method. The conclusions are drawn in Section 4.

% Motivated by the aforementioned analysis, we propose a fast and efficient HybridFormer under the attention-based encoder-decoder (AED) architecture to bridge this gap. The main contributions of this work are given as follows:

% 1) To overcome the quadratic complexity of the SA operator, a novel hybrid LASA paradigm based on rotary position embedding (RoPE\cite{num16su2021roformer}) is designed, i.e., RoPE-based SA and LA operators are incorporated in the squeeze and unsqueeze layers to improve the inference speed of the model. 

% 2) Second, we present a novel hybrid NAS-guided SRep mechanism which is called NSR and apply it to the convolution module to improve the performance of the model while not increasing any computational complexity during the inference phase. 

% 3) We contribute the first AED-based SqueezeFormer and the core code of our proposed HybridFormer on WeNet \cite{num22wu2021u2++} toolkit, which is available online.

% \vspace{-3mm}
% The remainder of this paper is organized as follows. In Section \ref{sec:METHODOLOGY}, we describe the details of the proposed method HybridFormer. In Section 3, the experimental dataset and setups are introduced to verify the effectiveness of the proposed method. The conclusions are drawn in Section 4.

% Many works \cite{num15tay2020efficient,num16su2021roformer,num17wang2020linformer,num18huang2019interlaced,num19qin2022cosformer,num20choromanski2020rethinking} have been devoted to the studies of LA to tackle this issue. 

\begin{figure*}[pt]
    \centering
    \includegraphics[height=6.5cm]{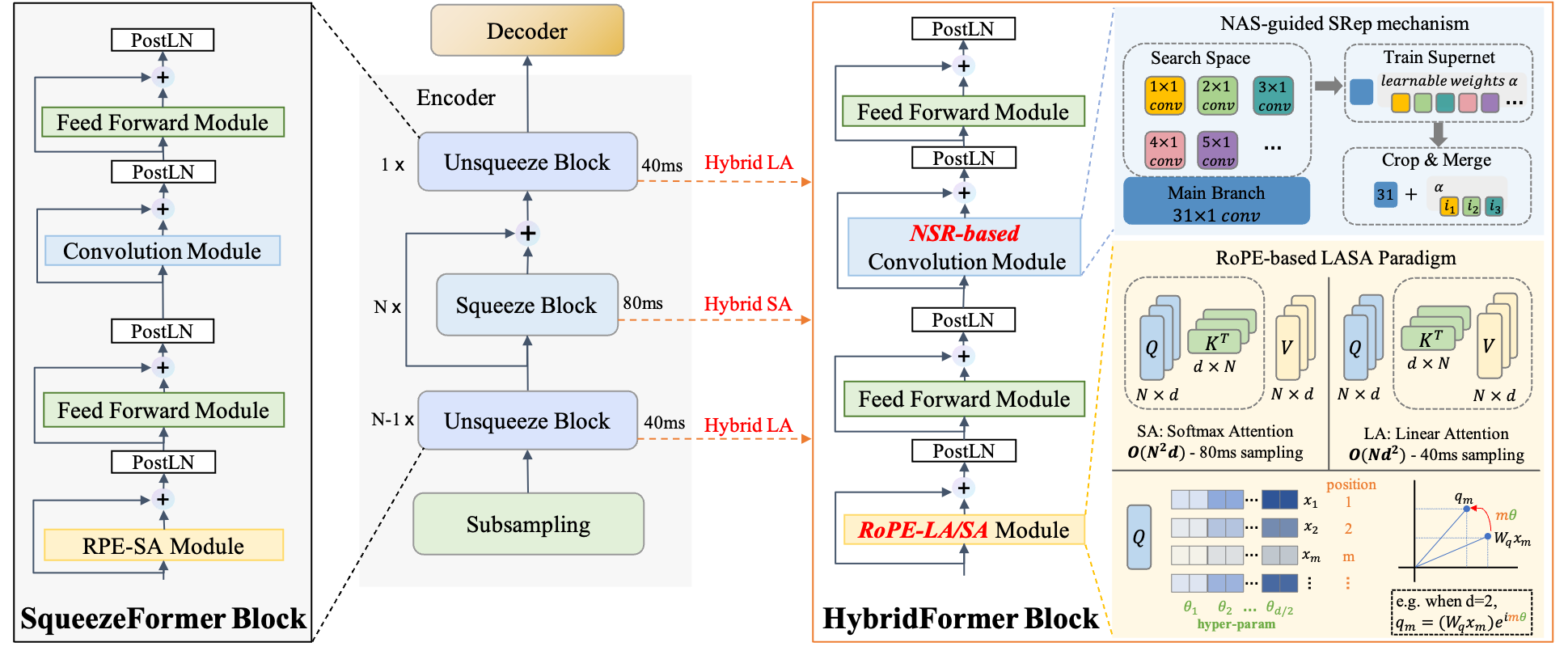}
    \vspace{-4mm}
    \caption{\label{fig:hybridformer} Overview of the SqueezeFormer (Left) and the proposed HybridFormer (Right).}
    \vspace{-4mm}
\end{figure*}

\vspace{-3mm}
\section{PROPOSED METHODOLOGY}
\label{sec:METHODOLOGY}

% In this section, we present a fast and efficient ASR method HybridFormer. First, the proposed LASA paradigm is introduced. Second, the detailed information of our proposed NSR mechanism is given. The overall architecture of the proposed method is illustrated in the right part of Fig. 1. 

\subsection{Overview}
% As illustrated in the right part of Fig. 1, the encoder of our proposed HybridFormer is mainly composed of three blocks, namely the subsampling block, Hybridformer LA block and Hybridformer SA block. In particular, the proposed hybrid NSR mechanism is adopted in all these blocks. 

% As illustrated in the right part of Fig. 1, the main components of the encoder part of HybridFormer are RoPE-based LA or SA module, feed forward module, NSR-based convolution module, and another feed forward module. 

% As illustrated in the right part of Fig. 1, RoPE-based LA and SA are adopted to unsqueeze blocks and squeeze blocks separately in the encoder to balance the performance and inference speed. Nevertheless, the proposed hybrid NSR mechanism is adopted in the convolution module of all blocks. 

As illustrated in the right part of Fig. 1, the encoder part of our proposed HybridFormer is mainly composed of subsampling block, HybridFormer LA block and HybridFormer SA block. In particular, the proposed hybrid NSR mechanism is adopted in all these blocks.

The decoder is the vanilla transformer which is not the main focus in this work. Besides the cross-entropy loss, connectionist temporal classification loss \cite{num23graves2006connectionist} is combined to jointly train our proposed method E2E.

\vspace{-3mm}
\subsection{Hybrid RoPE-based LASA Paradigm}

% To overcome the quadratic complexity with respect to the length of the input speech, SqueezeFormer reduced the sequence length of its squeeze blocks, as depicted in Fig. 1 (Left). 
To obtain faster inference speed, a natural idea is to use the LA operator to replace the SA operator. Nonetheless, according to the studies of \cite{num15tay2020efficient,num16su2021roformer,num17wang2020linformer,num18huang2019interlaced,num19qin2022cosformer,num20choromanski2020rethinking}, the performance of SA-based ASR models generally performs better than LA-based ASR models due to many factors, such as the approximation errors of low-rank decomposition or kernel-based LA methods. 

Hence, on account of the above-mentioned analysis, we present a novel hybrid LASA paradigm which is shown in Fig. \ref{fig:hybridformer} (Right), i.e., use LA and SA operators on different layers of the model upon its sampling rate to balance the inference speed and performance. Moreover, since the relative position embedding (RPE) used by SqueezeFormer is incompatible with the LA module, we adopt RoPE\cite{num16su2021roformer}-based SA and LA operators to mitigate this mismatch.

To be specific, assume $IS_N =\{x_i,i \in [1,N]\}$ denotes the N-length input sequence, where $x_i$ is the $i^{th}$ token of $IS_N$, and $d$ is the dimension. The proposed RoPE-based LASA paradigm first computes the corresponding query $q_m$ , key $k_n$, and value $v_n$ of the $x_m$ and $x_n$:
\begin{equation}
    \centering
    \begin{split}
        q_m &= f_q(x_m, m)=R^d_{\Theta, m} W_q x_m \\
        k_n &= f_k(x_n, n)=R^d_{\Theta, n} W_k x_n \\
        v_n &= f_v(x_n, n)=W_v x_n \\
    \end{split}
\end{equation}
\noindent where $x_m$ and $x_n$ are the $m^{th}$ and $n^{th}$ embeddings, $m$ and $n$ indicate the positions, $W_q$, $W_k$, and $W_v$ are the trainable matrices of the linear projection layer of $q_m$, $k_n$, and $v_n$, $\Theta = \{\theta_i = 10000^{-2(i-1)/d}, i \in [1, 2, \dots,d/2] \}$ is a set of pre-defined hyper-parameters,  $R^d_{\Theta, m} W_q x_n$ is a special case of $(x_m W_q)e^{im \theta}$ when $d=2$, where $\theta$ donates the angle of rotation here.
%, detailed derivation can be found in \cite{num16su2021roformer}. 
\begin{small}
\begin{equation}
    R^d_{\Theta,m}=
    \setlength{\arraycolsep}{1.2pt}
    \begin{pmatrix}
        \cos m\theta_1  & -\sin m\theta_1 & \cdots & 0 & 0 \\
        \sin m\theta_1 & \cos m\theta_1 & \cdots & 0 & 0 \\
        0 & 0 & \cdots & 0 & 0 \\
        \vdots & \vdots & \vdots & \cos m\theta_{d/2} & -\sin m\theta_{d/2} \\
        0 & 0 & \cdots & \sin m\theta_{d/2} & \cos m\theta_{d/2} \\
    \end{pmatrix}
\end{equation}
\end{small}
\noindent is the rotary matrix with $\Theta$. Then, the attention weight $A_{mn}$ is computed using $q_m$ and $k_n$:
\begin{equation}
    A_{mn} = q^{\mathrm{T}}_m k_n = x^{\mathrm{T}}_m W_q^{\mathrm{T}} R^d_{\Theta, n-m} W_k x_n
\end{equation}
\noindent where $R^d_{\Theta, n-m}=({R^d_{\Theta, m}})^T R^d_{\Theta, n}$. More detailed information can be referred to \cite{num16su2021roformer}.

When the sampling rate is 80ms, the RoPE-based SA operator is adopted, which is defined as:
\begin{equation}
    Hybrid_{SA}=\sum^{N}_{n=1} \frac{\text{exp}(A_{mn}/\sqrt{d})}{\sum^{N}_{j=1}\text{exp}(A_{mj}/\sqrt{d})} v_n
\end{equation}

When the sampling rate is 40ms which is relatively low, the RoPE-based LA operator is incorporated for acceleration, which is defined as:
% \begin{equation}
%     Hybrid_{LA}=\frac{\sum^{N}_{n=1}(R^d_{\Theta, m}\varphi (q_m))^{\mathrm{T}} (R^d_{\Theta, n}\varphi (k_n)) v_n} {\sum^{N}_{n=1} \varphi (q_m)^{\mathrm{T}} \varphi (k_n)}
% \end{equation}
\begin{equation}
    Hybrid_{LA}=\frac{\sum^{N}_{n=1}(R^d_{\Theta, m}\varphi (W_q x_m))^{\mathrm{T}} (R^d_{\Theta, n}\varphi (W_k x_n)) v_n} {\sum^{N}_{n=1} \varphi (W_q x_m)^{\mathrm{T}} \varphi (W_k x_n)}
\end{equation}
where  $\varphi(\cdot)$ is the non-negative function. 

\vspace{-2mm}
\subsection{NSR: Hybrid NAS-guided SRep Mechanism}

Despite the fact that SqueezeFormer is a convolution-enhanced transformer that could integrate the advantages of attention and convolution operations, we note that its convolution kernel size (i.e., 31) is relatively large, which weakens its capability to exploit local interactions to a certain extent. 

Consequently, we propose an effective hybrid NAS-guided SRep mechanism called NSR and apply it to the convolution module to address this issue. Noticeably, our NSR mechanism mainly consists of training and inference stages, among which the training stage includes search and re-train stages. The training phase aims to search for the most suitable convolution kernel combination for the current model using our proposed Darts [23] like technique, so as to improve the model's performance. The goal of the inference phase is to convert the multi-branch model into a single-branch model, so that no additional computational cost will be added during the inference phase.

\vspace{-2mm}
\subsubsection{Training Phase of our NSR-based Convolution Module}
\vspace{-1mm}

\begin{figure}[h]
\centering
    \vspace{-2mm}
    \setlength{\abovedisplayskip}{-8mm}
	\includegraphics[height=5.5cm]{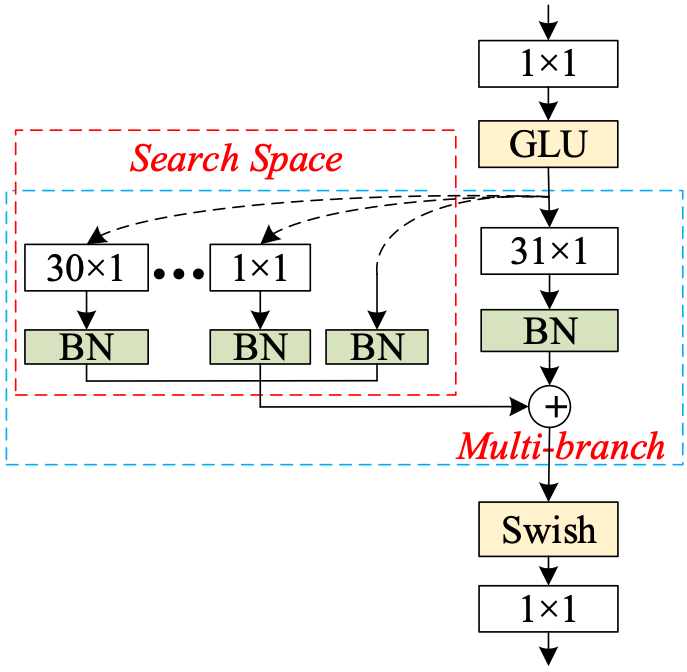}
	\caption{\label{fig:search} Training schematic of NSR-based convolution module}
    \setlength{\belowdisplayskip}{-3mm}
\end{figure}

As shown in Fig. \ref{fig:search}, the training phase of the proposed NSR-based convolution module starts with a gating mechanism \cite{num25dauphin2017language}, which is formed of a point-wise convolution layer and a gated linear unit (GLU) layer, then followed by our NSR-based depth-wise convolution layer which contains a trunk 31×1 convolution branch and additional convolution branches searched by our proposed NAS technique, a batchnorm (BN) layer, a Swish activation layer, and a point-wise convolution layer. 

\vspace{-1mm}
\vspace{-1mm}

% As for the NAS-searched branches, the search space is set up as a group of candidate operations $C=\{C^{(i,j)},  i \in [0,1,\dots,30],  j \in [0,1,\dots,11]\}$, where $i$ denotes the convolution operator with different kernel sizes (from 0 to 30, 0 denotes the branch with only BN layer) and $j$ denotes the $j^{th}$ block. To make the search space continuous, we adopt the operation of \cite{num26shi2021darts,num27liu2018darts}:

As for NAS-searched branches, a set of candidate operations $C=\{C^{(i,j)},  i \in [0,1,\dots,30],  j \in [0,1,\dots,11]\}$ is set up as the search space, where $C$ represents the Convolution with BN operations, $i$ represents the convolution operator with different kernel size (from 0 to 30, 0 represents the branch with only BN layer) and $j$ represents the $j^{th}$ block. To make the search space continuous, we adopt the operation of \cite{num26shi2021darts,num27liu2018darts}:
\begin{equation}
    \alpha_{(i,j)} = \frac{\text{exp}(c_{(i,j)})}{\sum^{30}_{k=0} \text{exp} (c_{k,j})}
\end{equation}

\noindent
where $\alpha (i,j)$ is the learnable weights parameterized by the vector $c (i,j)$. The main task of NAS is to capture the most suitable convolution operators by jointly training $\alpha$ and the model weights $w$, and its pseudo-code is described in Algorithm 1.

% \vspace{-6mm}
\begin{center}    
\vspace{-2mm}
\footnotesize
\begin{tabular}[t]{L{82.6mm}}
   \toprule
   \textbf{Algorithm 1}: \emph{\textbf{NSR}}-NAS guided SRep mechanism \\
   \midrule
   \textbf{Stage1 (search):} \\
    1. Create a mixed operation $C^{(i,j)}$ and the corresponding $\alpha^{i,j}$ \\
    2. Split the training dataset equally into train1 and train2 \\
    3. \textbf{while} warm-up \textbf{do}: \\
    \quad Update weight $w$ by descending $\nabla_{w} L_{train1} (w, \alpha)$ \\
    4. \textbf{while} not converged \textbf{do}: \\
    \quad 1) Update $\alpha$ by descending $\nabla_{\alpha} L_{train2} (w-\xi \nabla_w L_{train1} (w, \alpha), \alpha)$ \\
    \quad 2) Update $w$ by descending $\nabla_w L_{train1} (w, \alpha)$ \\
    5. Average the best checkpoints of $L_{val}$ with $\alpha$ \\
    \quad ($L_{train/val}$ stands for the loss on the train or valid dataset) \\
    \textbf{Stage2 (re-train):} \\
    1. Crop branches with negative learnable weights $\alpha^{(i,j)}$ \\
    2. Re-softmax the positive learnable weights as $\alpha^{\prime}$ \\
    3. Train the new model weight $w^\prime$ with the retained branches and $\alpha^{\prime}$ \\
    \quad by descending $\nabla_{w^\prime} L_{train} (w^\prime, \alpha^\prime)$\\
   \bottomrule
\end{tabular}
\end{center}

\vspace{-2mm}
\subsubsection{Inference Phase of our NSR-based Convolution Module}
\vspace{-1mm}

\begin{figure}[h]
\centering
    \vspace{-2mm}
    \setlength{\abovedisplayskip}{-5mm}
	\includegraphics[height=5.5cm]{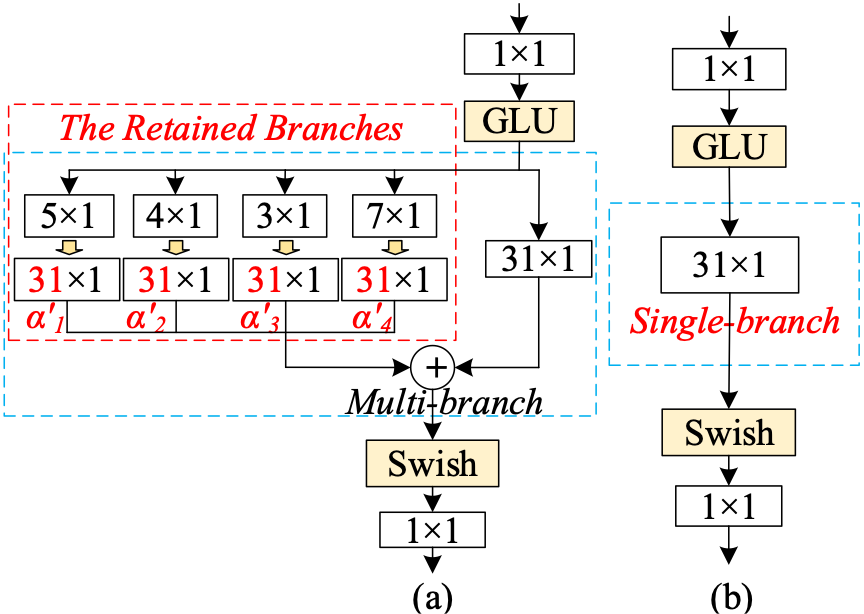}
	\caption{\label{fig:inference} Inference schematic of NSR-based convolution module}
    \setlength{\belowdisplayskip}{-5mm}
\end{figure}

As depicted in Fig. \ref{fig:inference}, the inference structure of the NSR-based convolution module will be converted into one single branch, and its corresponding parameters will be transformed as well. To be specific, the BN layer and convolution layer will be merged first. Then, the parameters of the NAS-searched and trunk convolution branches will also be merged, as the NAS-searched convolution branches can be considered as the degraded 31×1 convolution via simple algebraic transformation \cite{num28ding2021repvgg}. In this way, the NSR-based convolution module only has one single branch which is the 31×1 convolution layer during inference time. 

\vspace{-2mm}

\vspace{-2mm}
Accordingly, the ability of our model to exploit local features is greatly improved while not increasing any computational cost.
% Accordingly, our model's ability of exploiting local features is greatly improved while do not increase any computational cost.

\vspace{-1mm}
\section{EXPERIMENTS}
\label{sec:pagestyle}

\vspace{-1.5mm}
% \subsection{Dataset and Experimental Setups}
\subsection{Configurations}
\vspace{-1.5mm}

To verify the effectiveness of the proposed method, we conduct experiments on the public LibriSpeech \cite{num29panayotov2015librispeech} dataset. We train all models on the LibriSpeech training dataset, and evaluate them on LibriSpeech test-clean and test-other datasets. 

% To prove the validity of the proposed method, we conduct experiments on public LibriSpeech \cite{num29panayotov2015librispeech} dataset, which contains 1000 hours of 16kHZ labeled English speech and corresponding transcipts. We train all models on LibriSpeech training dataset, and evaluate them on LibriSpeech test-clean and test-other datasets. 

Specifically, audios are converted to 80-dimensional FBank. SentencePiece \cite{num30kudo2018sentencepiece} is used to construct a 5000 sub-word byte pair encoding tokenizer using the transcripts of LibriSpeech, and SpecAugment \cite{num31park2019specaugment} is used for data augmentation. All methods are optimized by Adamw \cite{num32loshchilov2017decoupled}, where the learning rate is 0.001, and the warmup ratio is 0.2 with the NoamAnnealing scheduler. All models contain 12 encoder blocks and 6 decoder blocks. As for the encoder, the convolution kernel size is 31, the number of attention heads is 4, the hidden dimension of the attention and FFN layers are 256 and 1024. As per the decoder, a single directional transformer with 4 attention heads is used, and the hidden dimension of the FFN layer is 2048. All experiments are conducted on WeNet and uniformly trained under 8 A100 GPUs.

% Specifically, the audios are converted to 80-dimensional FBank features, SentencePiece \cite{num30kudo2018sentencepiece} is used to construct a 5000 sub-word byte pair encoding tokenizer using the texts of LibriSpeech, and SpecAugment \cite{num31park2019specaugment} is used for data augmentation. All methods are optimized by Adamw \cite{num32loshchilov2017decoupled}, where the learning rate is 0.001, warmup ratio is 0.2, and hold ratio is 0.3 with NoamAnnealing scheduler. All models contain 12 encoder blocks and 6 decoder blocks. In terms of the encoder, the used convolution kernel size is 31, the number of attention heads is 4, the hidden dimension of the attention layer and FFN layer are 256 and 1024. As per the decoder, single directional transformer with 4 attention heads is used, and the hidden dimension of FFN layer is 2048. All experiments are conducted on WeNet toolkit and uniformly trained under 8 A100 GPUs.

\vspace{-3mm}
\subsection{Main Results}
\begin{figure}[h]
    \vspace{-2mm}
    \setlength{\abovedisplayskip}{-5mm}
	\centering
	\includegraphics[height=4.6cm]{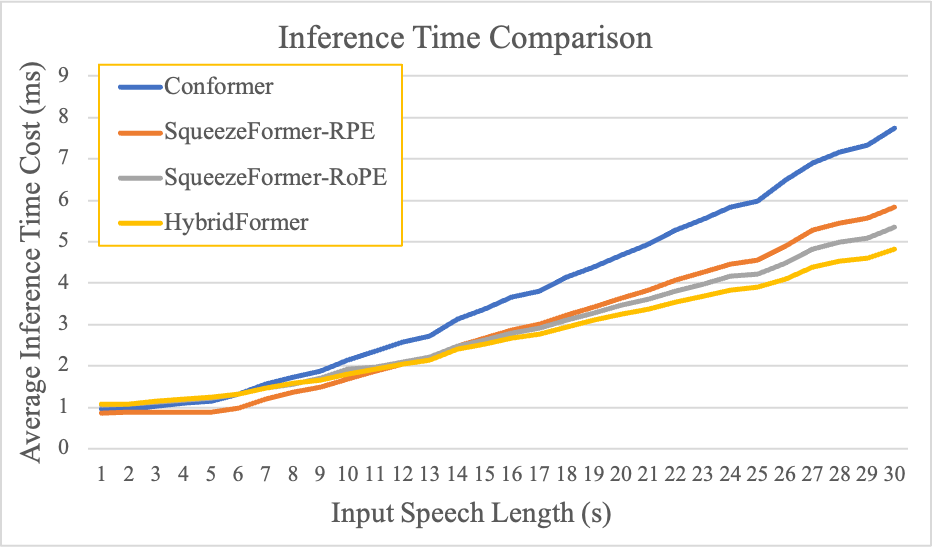}
        \vspace{-4mm}
	\caption{\label{fig:speed} Comparison of Inference speed on LibriSpeech.}
    \setlength{\belowdisplayskip}{-5mm}
    \vspace{-3mm}
\end{figure}

\vspace{-1mm}
In this section, we compare the WER and inference speed of our proposed HybridFormer with baseline SqueezeFormer and Conformer, which are illustrated in Table \ref{tb:all} and Fig. \ref{fig:speed}. In order to make a fair comparison, two evaluation methods CTC greedy search (CTC) and attention rescoring (Rescore) are adopted.

\begin{table}[htbp]\small
    \vspace{-5mm}
    \centering
    \caption{\label{tb:all} WER (\%) comparison of different models.}
    \begin{tabular}{cC{14mm}C{5mm}C{8.5mm}C{5mm}C{8.5mm}}
        \hline
        \toprule \\
        \specialrule{0em}{-14pt}{2pt}
        \multirow{2}{*}[-1mm]{Model} & \multirow{2}{14mm}[-1mm]{Params(M)} & \multicolumn{2}{c}{Test Clean} & \multicolumn{2}{c}{Test Other} \\
        \specialrule{0em}{0pt}{1pt}
        \cmidrule{3-6}  &  &  CTC & Rescore & CTC & Rescore   \\
        \specialrule{0em}{0pt}{-0.3pt}
        \midrule
        Conformer         & 22.15       & 3.82  & 3.28  & 9.77  & 8.86 \\
        \specialrule{0em}{0pt}{-0.3pt}
        Squeezeformer     & 22.22       & 3.51  & 3.07  & 9.28  & 8.44 \\
        \specialrule{0em}{0pt}{-0.3pt}
        HybridFormer      & \textbf{21.45}       & \textbf{3.34}  & \textbf{2.96}  & \textbf{8.44}  & \textbf{7.70} \\
        \specialrule{0em}{0pt}{-0.3pt}
        \bottomrule
    \end{tabular}
    \vspace{-2mm}
\end{table}

% In terms of the WER performance, it can be observed from Table \ref{tb:all} that our proposed HybridFormer consistently achieves the best WER with minimum parameters. Particularly, when we use CTC, our proposed method obtains a 4.8\% and a 9.1\% relative WER reduction on test-clean and test other datasets compared with baseline SqueezeFormer. When the Rescore is utilized, the proposed HybridFormer obtains a 3.6\% and an 8.8\% relative WER reduction on test-clean and test-other datasets.

In terms of the WER performance, it can be observed from Table \ref{tb:all} that our proposed HybridFormer consistently achieves the best WER with minimum parameters. Particularly, when we use CTC, our proposed method obtains a 4.8\% and a 9.1\% relative WER reduction on test-clean and test other datasets compared with baseline SqueezeFormer. When the Rescore is utilized, the proposed HybridFormer obtains a 3.6\% and an 8.8\% relative WER reduction on test-clean and test-other datasets.

\vspace{-0.5mm}

In terms of inference speed, all methods are tested on a single A10 GPU. From Fig. \ref{fig:speed}, we can see that our HybridFormer has the best performance. Compared with baseline SqueezeFormer, when the input speech is 30s length, the proposed HybridFormer increases the inference speed by relatively 18\%. 

\vspace{-2mm}
\subsection{Ablation Analysis}
In this section, we perform ablation experiments to prove the validity of the components of our proposed HybridFormer.

\vspace{-2mm}
\subsubsection{Effectiveness of the RoPE}
\vspace{-0.5mm}
First, the effectiveness of RoPE is evaluated.

\begin{table}[htbp]\small
    \vspace{-5mm}
    \centering
    \caption{\label{tb:rope} Effect of our proposed LASA paradigm.}
    \begin{tabular}{cC{5mm}C{8.5mm}C{5mm}C{8.5mm}}
        \hline
        \toprule \\
        \specialrule{0em}{-14pt}{2pt}
        \multirow{2}{*}[-1mm]{Model} & \multicolumn{2}{c}{Test Clean} & \multicolumn{2}{c}{Test Other} \\
        \specialrule{0em}{0pt}{-0.3pt}
        \cmidrule{2-5}  &  CTC & Rescore & CTC & Rescore   \\
        \specialrule{0em}{0pt}{-0.5pt}
        \midrule
        Squeezeformer       & \textbf{3.51}  & \textbf{3.07}  & 9.28  & 8.44 \\
        \specialrule{0em}{0pt}{-0.3pt}
         \quad + RoPE       & 3.67  & 3.14  & \textbf{9.10}  & \textbf{8.17} \\
         \specialrule{0em}{0pt}{-0.3pt}
        \bottomrule
    \end{tabular}
    \vspace{-3mm}
\end{table}

As can be observed from Table \ref{tb:rope}, when using CTC and Rescore, SqueezeFormer-RoPE performs better on the test-other dataset over vanilla SqueezeFormer: obtains a 1.9\% and a 3.2\% relative WER improvement but performs worse on the test-clean dataset: obtains a 4.6\% and a 2.3\% relative WER degradation.

\vspace{-3mm}
\subsubsection{Effectiveness of the proposed RoPE-LASA paradigm}

Then, we verify the effectiveness of the proposed hybrid RoPE-LASA paradigm.

\begin{table}[htbp]\small
    \vspace{-5mm}
    \centering
    \caption{\label{tb:LASA} Effect of our proposed LASA paradigm.}
    \vspace{1mm}
    \begin{tabular}{cC{5mm}C{8.5mm}C{5mm}C{8.5mm}}
        \hline
        \toprule \\
        \specialrule{0em}{-14pt}{2pt}
        \multirow{2}{*}[-1mm]{Model} & \multicolumn{2}{c}{Test Clean} & \multicolumn{2}{c}{Test Other} \\
        \specialrule{0em}{0pt}{-0.3pt}
        \cmidrule{2-5}  &  CTC & Rescore & CTC & Rescore   \\
        \specialrule{0em}{0pt}{1pt}
        \midrule
        \specialrule{0em}{0pt}{0.5pt}
        Squeezeformer-RoPE  & 3.67  & \textbf{3.14}  & 9.10  & 8.17 \\
        \specialrule{0em}{0pt}{-0.3pt}
         \quad + LASA       & \textbf{3.60}  & 3.16  & \textbf{8.98}  & \textbf{8.17} \\
         \specialrule{0em}{0pt}{-0.3pt}
        \bottomrule
    \end{tabular}
    \vspace{-2mm}
\end{table}

% \noindent
As can be seen from Table \ref{tb:LASA} and Fig. 4, compared with SqueezeFormer-RoPE, our proposed LASA paradigm can improve the inference speed while obtaining comparable WER. More, the acceleration effect of the proposed LASA-based model becomes more obvious as the input sequence gets longer.

\vspace{-2mm}
\subsubsection{Effectiveness of the proposed NSR mechanism}

Third, the effectiveness of our NSR mechanism is tested, and its ablation results are listed in Table \ref{tb:NSR}. 

\begin{table}[htbp]\small
    \vspace{-5mm}
    \centering
    \caption{\label{tb:NSR} Effect of our NSR mechanism.}
    \vspace{1mm}
    \begin{tabular}{cC{5mm}C{8.5mm}C{5mm}C{8.5mm}}
        \hline
        \toprule \\
        \specialrule{0em}{-14pt}{-0.5pt}
        \multirow{2}{*}[-0.7mm]{Model} & \multicolumn{2}{c}{Test Clean} & \multicolumn{2}{c}{Test Other} \\
        \specialrule{0em}{0pt}{-0.5pt}
        \cmidrule{2-5}  &  CTC & Rescore & CTC & Rescore   \\
        \specialrule{0em}{0pt}{-0.5pt}
        \midrule
        \specialrule{0em}{0pt}{0.5pt}
        Squeezeformer-RoPE  & 3.67  & 3.14  & 9.10  & 8.17 \\
        \specialrule{0em}{0pt}{0.5pt}
        \midrule
        \specialrule{0em}{0pt}{-0.1pt}
         \quad + SR-5       & \textbf{3.35}  & 2.93  & 8.56  & 7.78 \\
         \specialrule{0em}{0pt}{-0.3pt}
         \quad + SR-4       & 3.45  & 2.98  & 8.72  & 7.87 \\
         \specialrule{0em}{0pt}{-0.3pt}
         \quad + SR-3       & 3.40  & 3.04  & 8.71  & 7.90 \\
         \specialrule{0em}{0pt}{-0.3pt}
         \quad + SR-7       & 3.42  & 2.93  & 8.74  & 7.97 \\
         \specialrule{0em}{0pt}{-0.3pt}
        \midrule
        \specialrule{0em}{0pt}{-0.1pt}
         \quad \quad + \emph{\textbf{NSR}} & 3.37  & \textbf{2.92}  & \textbf{8.54}  & \textbf{7.65} \\
         \specialrule{0em}{0pt}{-0.3pt}
        \bottomrule
    \end{tabular}
\end{table}

\vspace{-2mm}

Here, numbers 5, 4, 3, and 7 denote the convolution kernel size of the retained branches, which are the top 4 branches searched by our proposed NAS. To be specific, the retained branches and the corresponding re-softmax weights $\alpha^\prime$ of the proposed NSR are [5, 4, 3, 7] and [0.377, 0.279, 0.246, 0.098] in our case.
% \noindent
% Table \ref{tb:NSR} lists the ablation results of our proposed NSR mechanism. Here, numbers 5, 4, 3, and 7 denote the convolution kernel size of the retained branches, which are the top 4 branches searched by our proposed NAS. To be specific, the retained branches and the corresponding re-softmax weights $\alpha^\prime$ of the proposed NSR are [5, 4, 3, 7] and [0.377, 0.279, 0.246, 0.098] in our case.

As illustrated in the table, we can observe that when just utilizing the SRep mechanism, the WER performance of the methods can be obviously improved compared with SqueezeFormer-RoPE, which is in accordance with our core idea. In addition, when using the proposed NSR mechanism, the WER performance of our method can be further improved by 6.2\% $\sim$ 8.2\% over SqueezeFormer-RoPE, which verifies the effectiveness of our proposed NSR mechanism.

% $\varphi (\cdot)$

\vspace{-1.5mm}
\section{CONCLUSIONS}
\label{sec:pagestyle}

% \vspace{-1mm}
In this paper, we propose a fast and efficient ASR method termed HybridFormer, which is a hybrid NSR-augmented LASA attention-convolution model. With our proposed hybrid LASA paradigm and NSR mechanism, HybridFormer could not only accelerate the inference speed but also enhance the model's feature representation ability, thus improving the recognition performance. Extensive experiments performed on the public LibriSpeech benchmark prove the effectiveness of our proposed method.

\vfill\pagebreak
\label{sec:refs}

\bibliographystyle{ieee}
% \bibliography{strings,refs}
% \setlength{\bibsep}{5pt}
\bibliography{hybridformer2}

\begin{thebibliography}{10}
\providecommand{\url}[1]{#1}
\csname url@samestyle\endcsname
\providecommand{\newblock}{\relax}
\providecommand{\bibinfo}[2]{#2}
\providecommand{\BIBentrySTDinterwordspacing}{\spaceskip=0pt\relax}
\providecommand{\BIBentryALTinterwordstretchfactor}{4}
\providecommand{\BIBentryALTinterwordspacing}{\spaceskip=\fontdimen2\font plus
\BIBentryALTinterwordstretchfactor\fontdimen3\font minus
  \fontdimen4\font\relax}
\providecommand{\BIBforeignlanguage}[2]{{%
\expandafter\ifx\csname l@#1\endcsname\relax
\typeout{** WARNING: IEEEtran.bst: No hyphenation pattern has been}%
\typeout{** loaded for the language `#1'. Using the pattern for}%
\typeout{** the default language instead.}%
\else
\language=\csname l@#1\endcsname
\fi
#2}}
\providecommand{\BIBdecl}{\relax}
\BIBdecl

\bibitem{num1vaswani2017attention}
A.~Vaswani, N.~Shazeer, N.~Parmar\emph{,~et~al.}, ``Attention is all you
  need,'' \emph{Advances in neural information processing systems}, vol.~30,
  2017.

\bibitem{num3zhang2020transformer}
Q.~Zhang, H.~Lu, H.~Sak\emph{,~et~al.}, ``Transformer transducer: A streamable
  speech recognition model with transformer encoders and rnn-t loss,'' in
  \emph{ICASSP 2020-2020 IEEE International Conference on Acoustics, Speech and
  Signal Processing (ICASSP)}.\hskip 1em plus 0.5em minus 0.4em\relax IEEE,
  2020, pp. 7829--7833.

\bibitem{num4miao2020transformer}
H.~Miao, G.~Cheng, C.~Gao\emph{,~et~al.}, ``Transformer-based online
  ctc/attention end-to-end speech recognition architecture,'' in \emph{ICASSP
  2020-2020 IEEE International Conference on Acoustics, Speech and Signal
  Processing (ICASSP)}.\hskip 1em plus 0.5em minus 0.4em\relax IEEE, 2020, pp.
  6084--6088.

\bibitem{num6gao2022paraformer}
Z.~Gao, S.~Zhang, I.~McLoughlin\emph{,~et~al.}, ``Paraformer: Fast and accurate
  parallel transformer for non-autoregressive end-to-end speech recognition,''
  \emph{arXiv preprint arXiv:2206.08317}, 2022.

\bibitem{num7li2019jasper}
J.~Li, V.~Lavrukhin, B.~Ginsburg\emph{,~et~al.}, ``Jasper: An end-to-end
  convolutional neural acoustic model,'' \emph{arXiv preprint
  arXiv:1904.03288}, 2019.

\bibitem{num8kriman2020quartznet}
S.~Kriman, S.~Beliaev, B.~Ginsburg\emph{,~et~al.}, ``Quartznet: Deep automatic
  speech recognition with 1d time-channel separable convolutions,'' in
  \emph{ICASSP 2020-2020 IEEE International Conference on Acoustics, Speech and
  Signal Processing (ICASSP)}.\hskip 1em plus 0.5em minus 0.4em\relax IEEE,
  2020, pp. 6124--6128.

\bibitem{num10majumdar2021citrinet}
S.~Majumdar, J.~Balam, O.~Hrinchuk\emph{,~et~al.}, ``Citrinet: Closing the gap
  between non-autoregressive and autoregressive end-to-end models for automatic
  speech recognition,'' \emph{arXiv preprint arXiv:2104.01721}, 2021.

\bibitem{num11gulati2020conformer}
A.~Gulati, J.~Qin, C.-C. Chiu\emph{,~et~al.}, ``Conformer:
  Convolution-augmented transformer for speech recognition,'' \emph{arXiv
  preprint arXiv:2005.08100}, 2020.

\bibitem{num12li2022ead}
C.~Li, Y.~Wang, F.~Deng\emph{,~et~al.}, ``Ead-conformer: a conformer-based
  encoder-attention-decoder-network for multi-task audio source separation,''
  in \emph{ICASSP 2022-2022 IEEE International Conference on Acoustics, Speech
  and Signal Processing (ICASSP)}.\hskip 1em plus 0.5em minus 0.4em\relax IEEE,
  2022, pp. 521--525.

\bibitem{num13narayanan2021cross}
A.~Narayanan, C.-C. Chiu, T.~O'Malley\emph{,~et~al.}, ``Cross-attention
  conformer for context modeling in speech enhancement for asr,'' in \emph{2021
  IEEE Automatic Speech Recognition and Understanding Workshop (ASRU)}.\hskip
  1em plus 0.5em minus 0.4em\relax IEEE, 2021, pp. 312--319.

\bibitem{num14chen2021continuous}
S.~Chen, Y.~Wu, Z.~Chen\emph{,~et~al.}, ``Continuous speech separation with
  conformer,'' in \emph{ICASSP 2021-2021 IEEE International Conference on
  Acoustics, Speech and Signal Processing (ICASSP)}.\hskip 1em plus 0.5em minus
  0.4em\relax IEEE, 2021, pp. 5749--5753.

\bibitem{num21kim2022squeezeformer}
S.~Kim, A.~Gholami, A.~Shaw\emph{,~et~al.}, ``Squeezeformer: An efficient
  transformer for automatic speech recognition,'' \emph{arXiv preprint
  arXiv:2206.00888}, 2022.

\bibitem{num15tay2020efficient}
Y.~Tay, M.~Dehghani, D.~Bahri\emph{,~et~al.}, ``Efficient transformers: A
  survey,'' \emph{ACM Computing Surveys (CSUR)}, 2020.

\bibitem{num16su2021roformer}
J.~Su, Y.~Lu, S.~Pan\emph{,~et~al.}, ``Roformer: Enhanced transformer with
  rotary position embedding,'' \emph{arXiv preprint arXiv:2104.09864}, 2021.

\bibitem{num17wang2020linformer}
S.~Wang, B.~Z. Li, M.~Khabsa\emph{,~et~al.}, ``Linformer: Self-attention with
  linear complexity,'' \emph{arXiv preprint arXiv:2006.04768}, 2020.

\bibitem{num18huang2019interlaced}
L.~Huang, Y.~Yuan, J.~Guo\emph{,~et~al.}, ``Interlaced sparse self-attention
  for semantic segmentation,'' \emph{arXiv preprint arXiv:1907.12273}, 2019.

\bibitem{num19qin2022cosformer}
Z.~Qin, W.~Sun, H.~Deng\emph{,~et~al.}, ``cosformer: Rethinking softmax in
  attention,'' \emph{arXiv preprint arXiv:2202.08791}, 2022.

\bibitem{num20choromanski2020rethinking}
K.~Choromanski, V.~Likhosherstov, D.~Dohan\emph{,~et~al.}, ``Rethinking
  attention with performers,'' \emph{arXiv preprint arXiv:2009.14794}, 2020.

\bibitem{aed}
J.~Chorowski, D.~Bahdanau, K.~Cho\emph{,~et~al.}, ``End-to-end continuous
  speech recognition using attention-based recurrent nn: First results,''
  \emph{arXiv preprint arXiv:1412.1602}, 2014.

\bibitem{num22wu2021u2++}
D.~Wu, B.~Zhang, C.~Yang\emph{,~et~al.}, ``U2++: Unified two-pass bidirectional
  end-to-end model for speech recognition,'' \emph{arXiv preprint
  arXiv:2106.05642}, 2021.

\bibitem{num23graves2006connectionist}
A.~Graves, S.~Fern{\'a}ndez, F.~Gomez\emph{,~et~al.}, ``Connectionist temporal
  classification: labelling unsegmented sequence data with recurrent neural
  networks,'' in \emph{Proceedings of the 23rd international conference on
  Machine learning}, 2006, pp. 369--376.

\bibitem{num25dauphin2017language}
Y.~N. Dauphin, A.~Fan, M.~Auli\emph{,~et~al.}, ``Language modeling with gated
  convolutional networks,'' in \emph{International conference on machine
  learning}.\hskip 1em plus 0.5em minus 0.4em\relax PMLR, 2017, pp. 933--941.

\bibitem{num26shi2021darts}
X.~Shi, P.~Zhou, W.~Chen\emph{,~et~al.}, ``Darts-conformer: Towards efficient
  gradient-based neural architecture search for end-to-end asr,'' \emph{arXiv
  preprint arXiv:2104.02868}, 2021.

\bibitem{num27liu2018darts}
H.~Liu, K.~Simonyan, and Y.~Yang, ``Darts: Differentiable architecture
  search,'' \emph{arXiv preprint arXiv:1806.09055}, 2018.

\bibitem{num28ding2021repvgg}
X.~Ding, X.~Zhang, N.~Ma\emph{,~et~al.}, ``Repvgg: Making vgg-style convnets
  great again,'' in \emph{Proceedings of the IEEE/CVF Conference on Computer
  Vision and Pattern Recognition}, 2021, pp. 13\,733--13\,742.

\bibitem{num29panayotov2015librispeech}
V.~Panayotov, G.~Chen, D.~Povey\emph{,~et~al.}, ``Librispeech: an asr corpus
  based on public domain audio books,'' in \emph{2015 IEEE international
  conference on acoustics, speech and signal processing (ICASSP)}.\hskip 1em
  plus 0.5em minus 0.4em\relax IEEE, 2015, pp. 5206--5210.

\bibitem{num30kudo2018sentencepiece}
T.~Kudo and J.~Richardson, ``Sentencepiece: A simple and language independent
  subword tokenizer and detokenizer for neural text processing,'' \emph{arXiv
  preprint arXiv:1808.06226}, 2018.

\bibitem{num31park2019specaugment}
D.~S. Park, W.~Chan, Y.~Zhang\emph{,~et~al.}, ``Specaugment: A simple data
  augmentation method for automatic speech recognition,'' \emph{arXiv preprint
  arXiv:1904.08779}, 2019.

\bibitem{num32loshchilov2017decoupled}
I.~Loshchilov and F.~Hutter, ``Decoupled weight decay regularization,''
  \emph{arXiv preprint arXiv:1711.05101}, 2017.

\end{thebibliography}

\end{document}